\documentclass[3p,twocolumn]{elsarticle}

\usepackage{amssymb} 
\usepackage{lineno}
\usepackage{epstopdf}
\usepackage{epsfig}
\usepackage{verbatim} 
\usepackage{rotating}
\usepackage{paralist}
\usepackage{atlasphysics}
\usepackage{subfigure}
\usepackage[labelfont=bf]{caption}
\usepackage{multirow} 
\usepackage{graphicx}
\usepackage{dcolumn}
\usepackage{bm}
\usepackage{hyperref}
\usepackage{amsmath} 
\usepackage{natbib}
\biboptions{numbers,sort&compress}

\newcommand{\beq}{\begin{equation}}
\newcommand{\eeq}{\end{equation}}
\newcommand{\bea}{\begin{eqnarray}}
\newcommand{\eea}{\end{eqnarray}}

\def\progname#1{{\sc #1}}

\newcommand\HfourL{\ensuremath{{H \rightarrow ZZ ^{*}\rightarrow 4\ell}}}

\newcommand\dSigFidi{\ensuremath{\frac{\mathrm{d} \sigma_{\mathrm{fid},i}}{\mathrm{d}x_i} }}

\newcommand\Nsigi{\ensuremath{n^{\mathrm{sig}}_i}}

\newcommand{\Lumi}{\ensuremath{\mathcal{L}_{\mathrm{int}}}}

\def\Powheg{\textsc{Powheg}}
\def\Minlo{\textsc{Minlo}}
\def\HResTwo{\textsc{HRes2}}

\def\ptH{\ensuremath{p_{\mathrm{T},H}}}
\def\yH{\ensuremath{|y_{H}|}}
\def\costhetastar{\ensuremath{|\cos{\theta^*}|}}
\def\mthreefour{\ensuremath{m_{34}}}
\def\njets{\ensuremath{n_{\mathrm{jets}}}}
\def\ptjet{\ensuremath{p_\mathrm{T,jet}}}

\def\ptHreco{\ensuremath{p^\mathrm{reco}_{\mathrm{T},H}}}
\def\yHreco{\ensuremath{|y^\mathrm{reco}_{H}|}}
\def\costhetastarreco{\ensuremath{|\cos{\theta^{*\mathrm{reco}}}|}}
\def\mthreefourreco{\ensuremath{m^\mathrm{reco}_{34}}}
\def\njetsreco{\ensuremath{n^\mathrm{reco}_{\mathrm{jets}}}}
\def\ptjetreco{\ensuremath{p_\mathrm{T,jet}^{\mathrm{reco}}}}

\def\ZZ{\ensuremath{ZZ}}
\def\WZ{\ensuremath{WZ}}
\def\Zpjets{\ensuremath{Z+\mathrm{jets}}}

\def\Z{\ensuremath{Z}}

\newcolumntype{d}[1]{D{.}{.}{#1}}

\newcommand{\papertitle}{
Fiducial and differential cross sections 
of Higgs boson production 
measured in the four-lepton decay channel 
in $\boldsymbol{pp}$ collisions at $\boldsymbol{\sqrt{s}}$~=~8~TeV   
with the ATLAS detector
}

\usepackage{preprintcover} 
\PreprintCoverPaperTitle{\papertitle}  
\PreprintIdNumber{CERN-PH-EP-2014-186}  
\PreprintCoverAbstract{Measurements of fiducial and differential cross sections of Higgs boson production in the ${H \rightarrow ZZ ^{*}\rightarrow 4\ell}$ decay channel are presented. 
The cross sections are determined within a fiducial phase space and corrected for detection efficiency and resolution effects. 
They are based on 20.3~fb$^{-1}$ of $pp$ collision data, produced at $\sqrt{s}$~=~8~TeV centre-of-mass energy at the LHC and recorded by the ATLAS detector. 
The differential measurements are performed in bins of transverse momentum and rapidity of the four-lepton system, the invariant mass of the subleading lepton pair 
and the decay angle of the leading lepton pair with respect to the beam line in the four-lepton rest frame, 
 as well as the number of jets and the transverse momentum of the leading jet. 
The measured cross sections are compared to selected theoretical calculations of the Standard Model expectations.
No significant deviation from any of the tested predictions is found.}  
\PreprintJournalName{PLB}

\clearpage

\begin{document}
\begin{frontmatter}
\title{
Fiducial and differential cross sections 
of Higgs boson production 
measured in the four-lepton decay channel 
in $pp$ collisions at $\sqrt{s}= 8$~TeV   
with the ATLAS detector
}
\author{The ATLAS Collaboration}
\address{}

\date{\today}

\begin{abstract}
Measurements of fiducial and differential cross sections of Higgs boson production in the ${H \rightarrow ZZ ^{*}\rightarrow 4\ell}$ decay channel are presented. 
The cross sections are determined within a fiducial phase space and corrected for detection efficiency and resolution effects. 
They are based on 20.3~fb$^{-1}$ of $pp$ collision data, produced at $\sqrt{s}$~=~8~TeV centre-of-mass energy at the LHC and recorded by the ATLAS detector. 
The differential measurements are performed in bins of transverse momentum and rapidity of the four-lepton system, the invariant mass of the subleading lepton pair 
and the decay angle of the leading lepton pair with respect to the beam line in the four-lepton rest frame, 
 as well as the number of jets and the transverse momentum of the leading jet. 
The measured cross sections are compared to selected theoretical calculations of the Standard Model expectations.
No significant deviation from any of the tested predictions is found.
\end{abstract}

\end{frontmatter}

\section{Introduction}
\label{sec:Introduction}
In 2012 the ATLAS and CMS collaborations announced the discovery of a new
particle~\cite{ATLAS:2012af,Chatrchyan:2012ia} in the search for the Standard Model (SM) Higgs boson~\cite{Englert:1964et,Higgs:1964ia,Higgs:1964pj,Guralnik:1964eu,Higgs:1966ev,Kibble:1967sv} at
the CERN Large Hadron Collider (LHC)~\cite{lhc}. 
Since this discovery, the particle's mass $m_H$ was measured by the ATLAS and CMS collaborations~\cite{CombinedMass,CMSfourl,CMSyy}. The result of the ATLAS measurement based on 25 \ifb\ 
of data collected at centre-of-mass energies of 7 TeV and 8 TeV is $125.36 \pm 0.41$~GeV.
Tests of the couplings and spin/CP quantum numbers have been reported by both collaborations~\cite{CMSfourl,ATLAS:Spin,Aad:2013wqa} and show agreement with the predicted scalar nature 
of the SM Higgs boson. 
 
In this Letter, measurements of fiducial and differential production cross sections for the \HfourL\ decay channel are 
reported and compared 
to selected theoretical calculations. 
The event selection and the background determination are the same as in Ref.~\cite{MassMuZZ}, where a detailed description is given.
For this measurement, an integrated luminosity of 20.3 \ifb\ of $pp$ collisions is analysed. The data were collected at the LHC at a centre-of-mass energy of $\sqrt{s}$~=~8~TeV and 
recorded with the ATLAS detector~\cite{Aad:2008zzm}. 
 
The ATLAS detector covers the pseudorapidity range ${ | \eta | < 4.9}$ and the full azimuthal
angle $\phi$.\footnote{ATLAS uses a right-handed coordinate system with its origin
at the nominal interaction point (IP) at the centre of the detector and the $z$-axis along the beam
pipe. The $x$-axis points from the IP to the centre of the LHC ring, and the $y$-axis points
upward. Cylindrical coordinates $(r,\phi)$ are used in the transverse plane, $\phi$ being the
azimuthal angle around the beam pipe. The pseudorapidity is defined in terms of the polar angle
$\theta$ as ${\eta=-\ln[\tan(\theta/2)]}$.} 
It consists of an inner tracking detector covering the pseudorapidity range  ${ | \eta | < 2.5}$ surrounded
 by a superconducting solenoid, electromagnetic and hadronic calorimeters, and an external muon spectrometer 
with large superconducting toroidal magnets.

Fiducial cross sections are quoted to minimize the model dependence of the acceptance corrections related to the 
extrapolation to phase-space regions not covered by the detector.
The measured fiducial cross sections are corrected for detector effects to be directly compared to theoretical calculations.

The differential measurements are performed in several observables related to the Higgs boson production and decay. 
These include the transverse momentum \ptH\ and rapidity \yH\ of the Higgs boson, the invariant mass of the subleading lepton pair \mthreefour\ (the leading and subleading lepton pairs are defined in Section~\ref{sec:Selection}) 
and the magnitude of the cosine of the decay angle of the leading lepton pair in the four-lepton rest frame with respect to the beam axis \costhetastar. 
The number of jets \njets\ and the transverse momentum of the leading jet \ptjet\ are also included. 
The distribution of the \ptH\ observable is sensitive to the Higgs boson production mechanisms as well as spin/CP quantum numbers, and can be used to test perturbative QCD predictions.
This distribution has been studied extensively and precise predictions exist (see e.g. Refs.~\cite{Bozzi:2005wk,deFlorian:2011xf,HRes,HRes2,LHCHiggsCrossSectionWorkingGroup:2012vm}), 
including the effect of finite quark masses. 
The distribution of the \yH\ observable can be used to probe the parton distribution functions (PDFs) of the proton. 
The distributions of the decay variables \mthreefour\ and \costhetastar\ are sensitive to the Lagrangian structure of Higgs boson interactions, e.g.\ spin/CP quantum numbers 
and higher-dimensional operators.
The jet multiplicity and transverse momentum distributions are sensitive to QCD radiation effects and to the relative rates of Higgs boson production modes. 
The distribution of the transverse momentum of the leading jet probes quark and gluon radiation.

\section{Theoretical predictions and simulated samples}
\label{sec:Samples}
The Higgs boson production cross sections and decay branching fractions as well as their
uncertainties are taken from
Refs.~\cite{LHCHiggsCrossSectionWorkingGroup:2012vm,LHCHiggsCrossSectionWorkingGroup:2011ti}. The
cross sections for the gluon-fusion (ggF) process have been calculated to
next-to-leading order (NLO)~\cite{Djouadi:1991tka,Dawson:1990zj,Spira:1995rr}, and 
next-to-next-to-leading order (NNLO)~\cite{Harlander:2002wh,Anastasiou:2002yz,Ravindran:2003um}
 in QCD with additional next-to-next-to-leading logarithm (NNLL) soft-gluon resummation~\cite{Catani:2003zt}.
The cross section values have been modified to include NLO electroweak (EW) radiative corrections, assuming factorization between QCD and EW effects~\cite{Aglietti:2004nj,Actis:2008ug,deFlorian:2012yg,Anastasiou:2012hx,Baglio:2010ae}.  
The cross sections for the vector-boson fusion (VBF) processes are calculated with full NLO QCD and EW
corrections~\cite{Ciccolini:2007jr,Ciccolini:2007ec,Arnold:2008rz},
 and approximate NNLO QCD
corrections are included~\cite{Bolzoni:2010xr}.
  The cross sections for the associated $WH/ZH$
production processes ($VH$) are calculated at NLO~\cite{Han:1991ia} and at NNLO~\cite{Brein:2003wg} 
 in QCD,
and NLO EW radiative corrections~\cite{Ciccolini:2003jy}
 are applied. 
 The cross sections for
associated Higgs boson production with a $t\bar{t}$ pair ($t\bar{t}H$) are calculated at NLO
in QCD~\cite{Beenakker:2001rj,Beenakker:2002nc,Dawson:2002tg,Dawson:2003zu}.

The Higgs boson branching fractions for decays 
 to four-lepton final
states are provided by \progname{Prophecy4f}~\cite{Bredenstein:2006rh,Bredenstein:2006ha}, which
implements the complete NLO QCD+EW corrections and interference effects between identical final-state
fermions. 

The \HfourL\ signal is modelled using the \Powheg\ Monte Carlo (MC) event
generator~\cite{powheg1,powheg2,powheg3,powheg4,powheg5}, which calculates separately the ggF and
VBF production mechanisms with matrix elements up to NLO. 
The description of the Higgs boson transverse momentum spectrum in the ggF process is adjusted to follow the calculation in Ref.~\cite{HRes,HRes2}, which includes QCD corrections up to NLO and QCD soft-gluon resummations up to NNLL, as well as finite quark masses~\cite{Bagnaschi:2011tu}.  
\Powheg\ is interfaced to \progname{Pythia8}~\cite{pythia81} for showering and hadronization, which in turn is
interfaced to \progname{Photos}~\cite{Golonka:2005pn,Davidson:2010ew} to model photon radiation in the final state. \progname{Pythia8} is used to simulate $VH$ and $t\bar{t}H$ production.
The response of the ATLAS detector is modelled in a simulation~\cite{simuAtlas} based on GEANT4~\cite{GEANT4}.

The measured fiducial cross-section distributions are compared to three ggF theoretical calculations: \Powheg\ without the adjustments to the \ptH\ spectrum described above, \Powheg\ interfaced to \Minlo\ (Multi-scale improved NLO)~\cite{Minlo} 
and \HResTwo\ (v.2.2)~\cite{HRes,HRes2}. 
\Powheg\ with \Minlo\ provides predictions for jet-related variables at NLO for Higgs boson production in association with one jet.
The \HResTwo\ program computes fixed-order cross sections for ggF SM Higgs boson production up to NNLO. 
All-order resummation of soft-gluon effects at small transverse momenta is consistently included up to NNLL, using dynamic factorization and resummation scales.
The program implements top- and bottom-quark mass dependence up to NLL+NLO. 
At NNLL+NNLO level only the top-quark contribution is considered. \HResTwo\ does not perform showering and QED final-state radiation effects are not included. 

The contributions from the other production modes are added to the ggF predictions. 
At a centre-of-mass energy of 8 TeV and for a Higgs boson mass of 125.4 GeV, their relative contributions to the total cross 
section are 87.3\% (ggF), 
7.1\% (VBF), 3.1\% ($WH$), 1.9\% ($ZH$) and 0.6\% ($t\bar{t}H$), respectively.

All theoretical predictions are computed for a SM Higgs boson with mass 125.4~GeV.
They are normalized to the most precise SM inclusive cross-section predictions currently available~\cite{Heinemeyer:2013tqa}, corrected for the fiducial acceptance derived from the simulation.

The \ZZ, \WZ, \ttbar\ and \Zpjets\ background events are modelled using the simulated samples and cross sections described in Ref.~\cite{MassMuZZ}.

\section{Event selection}
\label{sec:Selection}
The detector level physics object definitions of muons, electrons, and jets, and the event selection applied in this analysis are the same as in Ref.~\cite{MassMuZZ}, with the exception of the jet selection and the 
additional requirement on the four-lepton invariant mass described below.
A brief overview is given in this section.

Events with at least four leptons are selected with single-lepton and dilepton triggers. 
The transverse momentum and transverse energy thresholds for the single-muon and single-electron triggers are 24~GeV. 
Two dimuon triggers are used, one with symmetric thresholds at 13~GeV and the other with asymmetric thresholds at 18~GeV and 8~GeV. 
For the dielectron trigger the symmetric thresholds are 12~GeV.
Furthermore there is an electron--muon trigger with thresholds at 12~GeV (electron) and 8~GeV (muon).

Higgs boson candidates are formed by selecting two same-flavour opposite-sign (SFOS) lepton pairs (a lepton
quadruplet).  
The leptons must satisfy identification, impact parameter, and track-based and calorimeter-based isolation criteria. Each muon (electron) must satisfy transverse momentum  $\pt>6$~GeV (transverse energy $\et>7$~GeV) and be in the pseudorapidity
range ${\left|\eta\right|<2.7}$ (2.47).  The highest-\pt~lepton in the quadruplet
must satisfy ${\pt >20}$~GeV, and the second (third) lepton in $\pt$ order must satisfy $\pt>15$~(10)~GeV. 
The leptons are required to be separated from each other by ${\Delta R \equiv \sqrt{(\Delta \eta)^2 + (\Delta \phi)^2} >}$~0.1 (0.2) when having the same (different) lepton flavours.  

Multiple quadruplets within a single event are possible: for four muons or four electrons there are two ways to pair the masses, and for five or more leptons there are multiple combinations. 
The quadruplet selection is done separately in each channel: $4\mu$, $2e2\mu$, $2\mu2e$, $4e$, keeping only a single
 quadruplet per channel. Here the first flavour index refers to the leading lepton pair, which is the pair with the invariant mass  $m_{12}$ closest to the $Z$ boson mass~\cite{PDG-2012}. 
The invariant mass $m_{12}$ is required to be between 50~GeV and 106~GeV. 
The subleading pair of each channel is chosen as the remaining pair with mass $m_{34}$ closest to the $Z$ boson mass and satisfying the requirement $12 < \mthreefour <  115$~GeV. 
Finally, if more than one channel has a quadruplet
passing the selection, the channel with the highest expected signal rate is kept, in the order:
$4\mu$, $2e2\mu$, $2\mu2e$, $4e$. A $J/\psi$~veto is applied: ${m(\ell_{i},\ell_{j}) > 5}$~GeV for SFOS lepton pairs. 
Only events with a four-lepton invariant mass in the range 118--129 GeV are kept. 
This requirement defines the signal mass window and was chosen by minimizing the expected uncertainty on the total signal yield determination, 
taking into account the experimental uncertainty on the Higgs boson mass.

Jets are reconstructed from topological clusters of calorimeter cells using the anti-$k_t$ algorithm~\cite{AntiKt} with the distance parameter $R=0.4$.
In this analysis, jets \cite{jetcalib} are selected by requiring ${\pt\ > 30}$~GeV, ${|y| < }$~4.4 and, in order to avoid double counting of electrons that are also reconstructed as jets, ${\Delta R (\mathrm{jet, electron}) >}$~0.2. 
 
The events are divided into bins of the variables of interest, which are computed with the reconstructed four-momenta of the selected lepton quadruplets or from the reconstructed jets: the transverse momentum \ptHreco\ and the rapidity \yHreco\ of the four-lepton system,  the invariant mass of the subleading lepton pair \mthreefourreco, the magnitude of the cosine of the decay angle of the leading lepton pair in the four-lepton rest frame with respect to the beam axis \costhetastarreco, the number of jets \njetsreco, and the transverse momentum of the leading jet \ptjetreco.
In order to distinguish them from the unfolded variables used in the cross section bin definition, they are labelled with ``reco''.

\section{Definition of the fiducial region}
\label{sec:Fiducial}
The fiducial selection, outlined in Table~\ref{tab:FiducialRegion}, is designed to replicate at simulation level, before applying detector effects, the analysis selection as closely as possible in order to minimize model-dependent acceptance effects on the measured cross sections.

\begin{table}
\centering
\caption{List of selection cuts which define the fiducial region of the cross section measurement. Same flavor opposite sign lepton pairs are denoted as SFOS, the leading lepton pair mass as $m_{12}$, and the subleading lepton pair mass as $m_{34}$. \label{tab:FiducialRegion} }
\vspace{0.2cm}
\begin{tabular}{ll}
\hline\hline
\multicolumn{2}{c}{\bf Lepton selection} \\
Muons: & $\pt>6 \GeV, |\eta|<2.7$  \\
Electrons: & $\pt>7 \GeV, |\eta|<2.47$  \\
\hline
\multicolumn{2}{c}{\bf Lepton pairing} \\
Leading pair: & SFOS lepton pair with \\
& smallest $|m_Z-m_{\ell\ell}|$  \\
Subleading pair: & Remaining SFOS \\
& lepton pair with \\
& smallest $|m_Z-m_{\ell\ell}|$  \\
\hline
\multicolumn{2}{c}{\bf Event selection} \\
Lepton kinematics: & $\pt > 20, 15, 10 $ \GeV  \\
Mass requirements: & $50 < m_{12} < 106$ \GeV\\
& $12 < m_{34} < 115$ \GeV  \\
Lepton separation: & $\Delta R(\ell_{i},\ell_{j}) > 0.1~(0.2)$ \\
& for same- (different-) \\
& flavour leptons \\
J/\psi~veto: & $m(\ell_{i},\ell_{j}) > 5$ \GeV \\ 
& for all SFOS lepton pairs  \\
Mass window: &  $118 < m_{4\ell} < 129 $ \GeV  \\
\hline\hline
\end{tabular}
\end{table}

The fiducial selection is applied to electrons and muons originating from vector-boson decays before they emit photon radiation, referred to as Born-level leptons.
An alternative approach would be to correct the lepton momenta by adding final-state radiation photons within a cone of size $\Delta R~<$~0.1 around each lepton (dressing).
For this analysis the acceptance difference between Born and dressed-lepton definitions is less than 0.5\%. Particle-level jets are reconstructed from all stable particles except muons and neutrinos using the anti-$k_t$ algorithm 
with the distance parameter $R$~=~0.4. 

Jets are selected by requiring \pt~$>$~30~GeV, $|y|$~$<$~4.4 and
${\Delta R\mathrm{(jet, electron)} > 0.2}$.
Muons (electrons) must satisfy \pt~$>$~6~(7)~GeV and $| \eta | <$~2.7~(2.47). 
Events  in which at least one of the \Z\ bosons decays into $\tau$ leptons are removed.
Quadruplets are formed from two pairs of SFOS leptons.  
The leptons are paired as in Section~\ref{sec:Selection}, including the possibility of incorrectly pairing the leptons, which happens in about 5\% of the selected events for a SM Higgs boson with mass 125.4~GeV. 
The leading pair is defined as the SFOS lepton pair with invariant mass $m_{12}$ closest to the \Z\ boson mass and the subleading pair is defined as the remaining SFOS 
lepton pair with invariant mass $m_{34}$ closest to the \Z\ boson mass.

The three highest-\pt\ leptons in the quadruplet are required to have \pt~$>$~20, 15, 10~GeV, respectively, and the lepton pairs must have 50~$<$~$m_{12}$~$<$~106~GeV and 12~$<$~$m_{34}$~$<$~115~GeV. 

The separation between the leptons is required to be ${\Delta R(\ell_{i},\ell_{j}) > 0.1~(0.2)}$ for same- \mbox{(different-)} flavour leptons. 
A $J/\psi$~veto is applied: ${m(\ell_{i},\ell_{j}) > 5}$~GeV for all SFOS lepton pairs. Furthermore, the mass of the four-lepton system $m_{4
\ell}$ must be close to $m_H$, i.e. ${118 < m_{4\ell} < 129}$~GeV.

For a SM Higgs boson mass of 125.4~GeV, the acceptance of the fiducial selection (with respect to the full phase space of $H \rightarrow\ ZZ^* \rightarrow 2\ell 2\ell'$, where $\ell,\ell' = e, \mu$) is 45.7\%. The number of events passing the event selection divided by the number of events passing the fiducial selection is 55.3\%; about 1\% of the events passing the event selection do not pass the fiducial selection.

\section{Background estimate}
\label{sec:Background}
The background estimates used in this analysis are described in detail in Ref.~\cite{MassMuZZ}.
The irreducible \ZZ\ and the reducible \WZ\ background contributions are estimated using simulated samples normalized to NLO predictions.
For the jet-related variables, the simulation predictions are compared to data for $m_{4\ell} > 190$ GeV where the \ZZ\ background process is dominant; 
shape differences between the distributions in data and simulation are used to estimate systematic uncertainties.

The reducible \Zpjets\ and \ttbar\ background contributions are estimated with data-driven methods. Their normalizations are obtained from data control regions and extrapolated to the signal region using transfer factors.  
The $\ell\ell+\mu\mu$ final state is dominated by $Z$ + heavy-flavour jets and the $\ell\ell+ee$ final state by $Z$ + light-flavour jets. The misidentification of light-flavour jets as electrons is difficult to model in the simulation. 
Therefore the distributions for $\ell\ell + ee$ are taken from data control regions and extrapolated to the signal region, while the background distributions for $\ell\ell+\mu\mu$ are taken from simulated samples. 

After the analysis selection about 9 background events are expected: 6.7 events from irreducible \ZZ\ and 2.2 events from the reducible background.

The observed distributions compared to the signal and background expectations for the six reconstructed observables \ptHreco, \yHreco, \mthreefourreco, \costhetastarreco, \njetsreco, and \ptjetreco\
are shown in Fig.~\ref{Fig:DiffObservedYieldsMain}. 
The signal prediction includes VBF, $ZH$, $WH$, $t\bar{t}H$, and the \Powheg\ ggF calculation for a Higgs boson with $m_H = 125$~GeV and is normalized to the most precise SM inclusive cross-section calculation currently available~\cite{Heinemeyer:2013tqa}.

\begin{figure*}[htbp]
   \begin{center}
      \subfigure[\label{Fig:DiffObservedYields:pt}]{\includegraphics[width=0.45\linewidth]{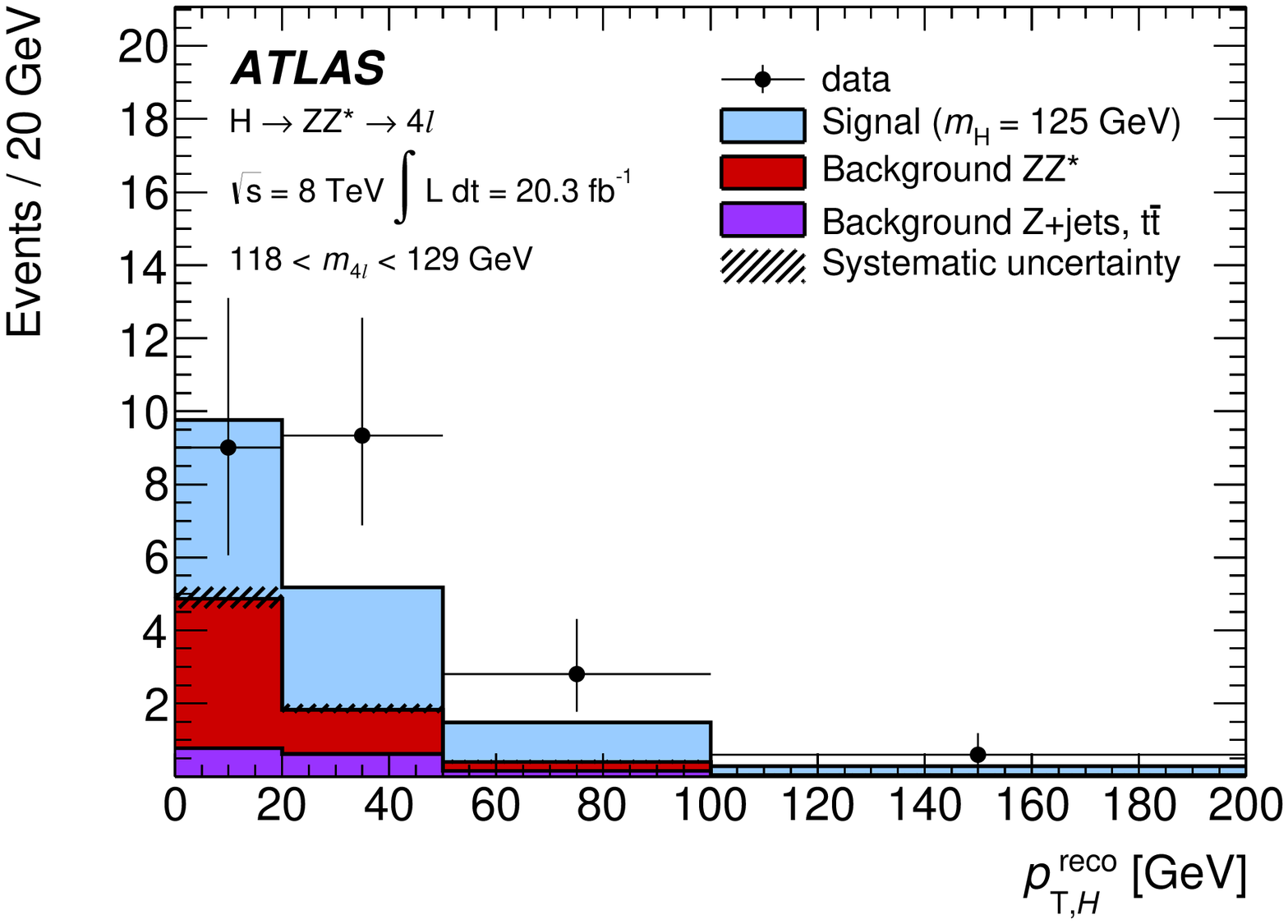}}
      \subfigure[\label{Fig:DiffObservedYields:rapidity}]{\includegraphics[width=0.45\linewidth]{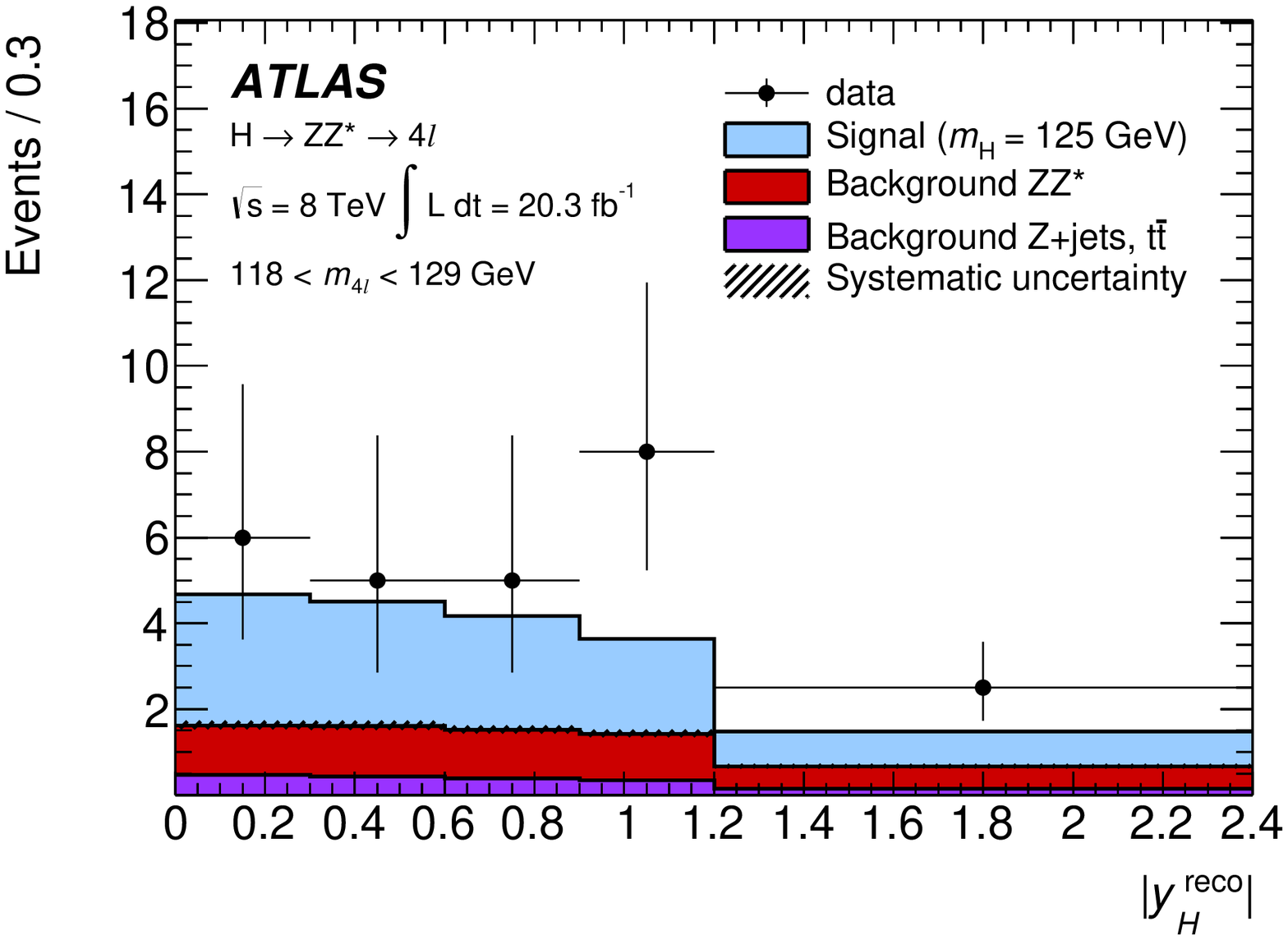}}
      \subfigure[\label{Fig:DiffObservedYields:m34}]{\includegraphics[width=0.45\linewidth]{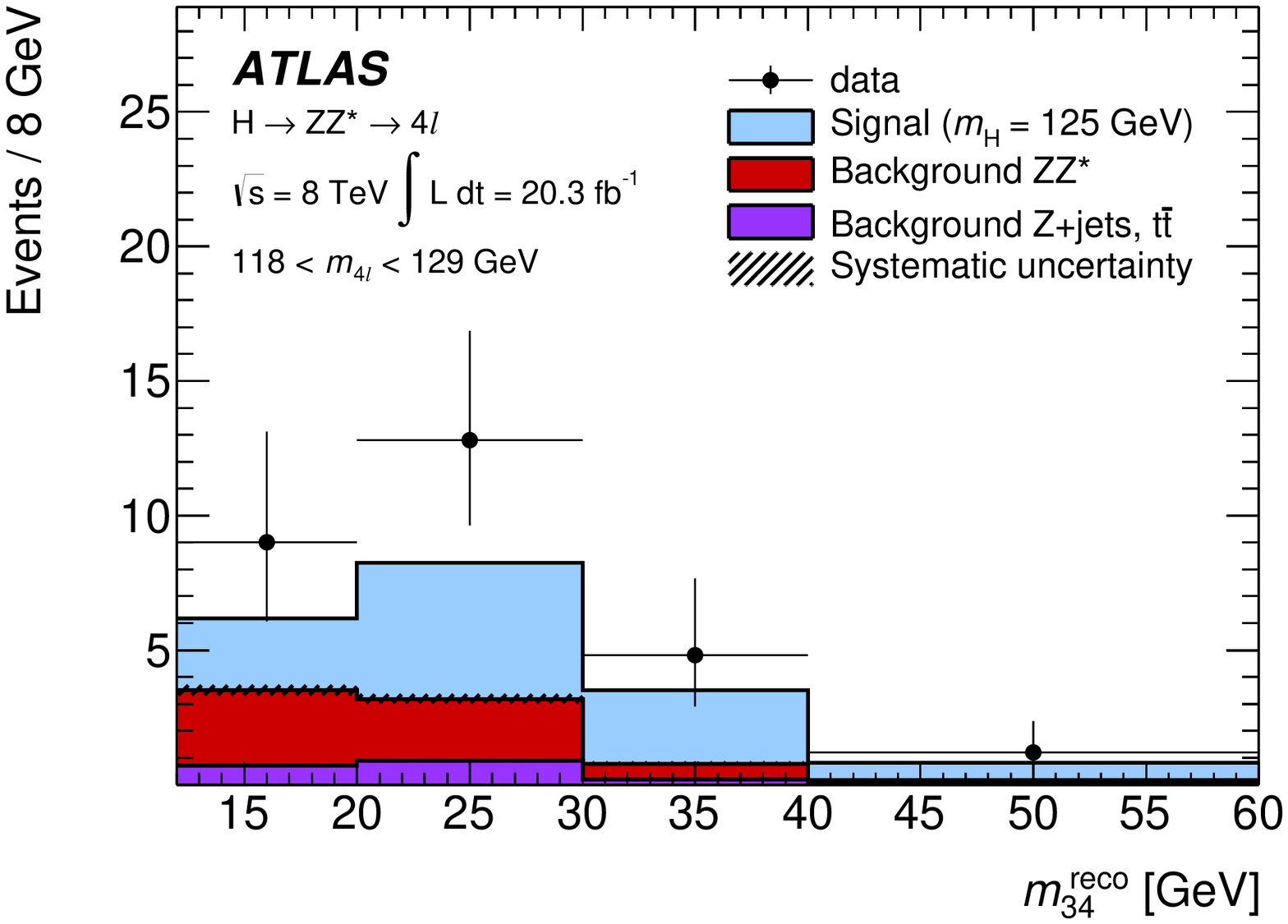}}
      \subfigure[\label{Fig:DiffObservedYields:costhetastar}]{\includegraphics[width=0.45\linewidth]{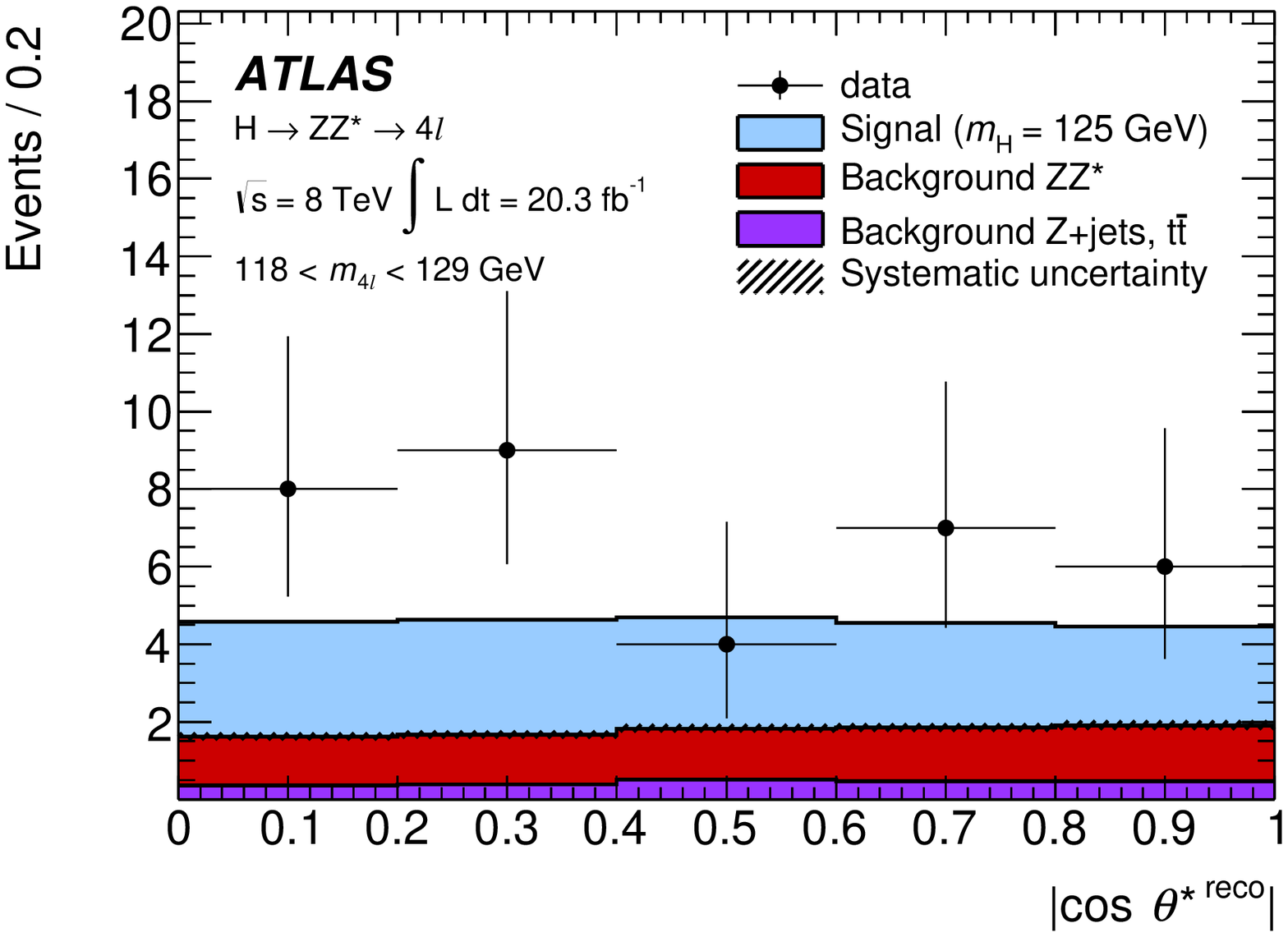}}
      \subfigure[\label{Fig:DiffObservedYields:njets}]{\includegraphics[width=0.45\linewidth]{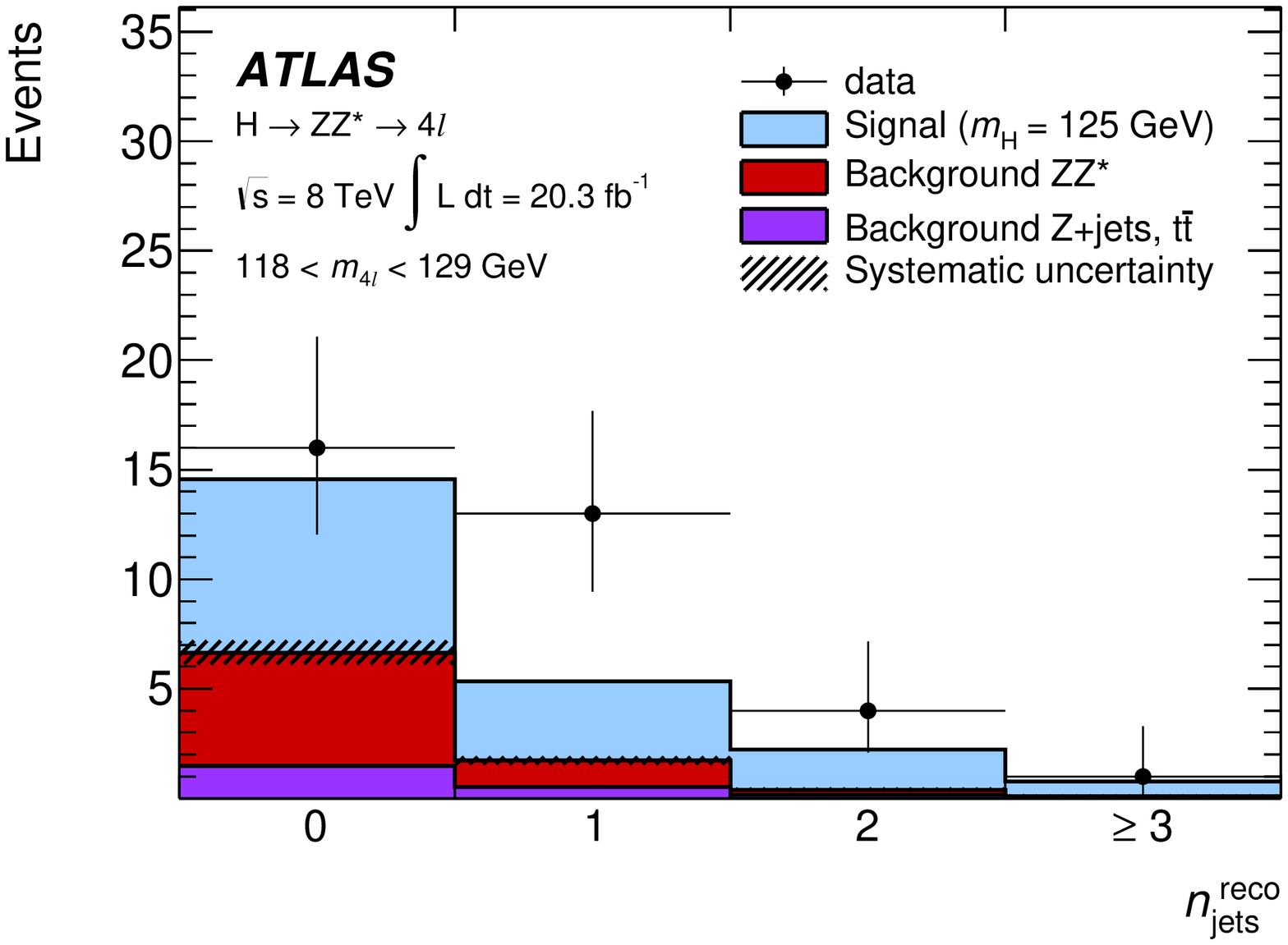}}
      \subfigure[\label{Fig:DiffObservedYields:leadingjetpt}]{\includegraphics[width=0.45\linewidth]{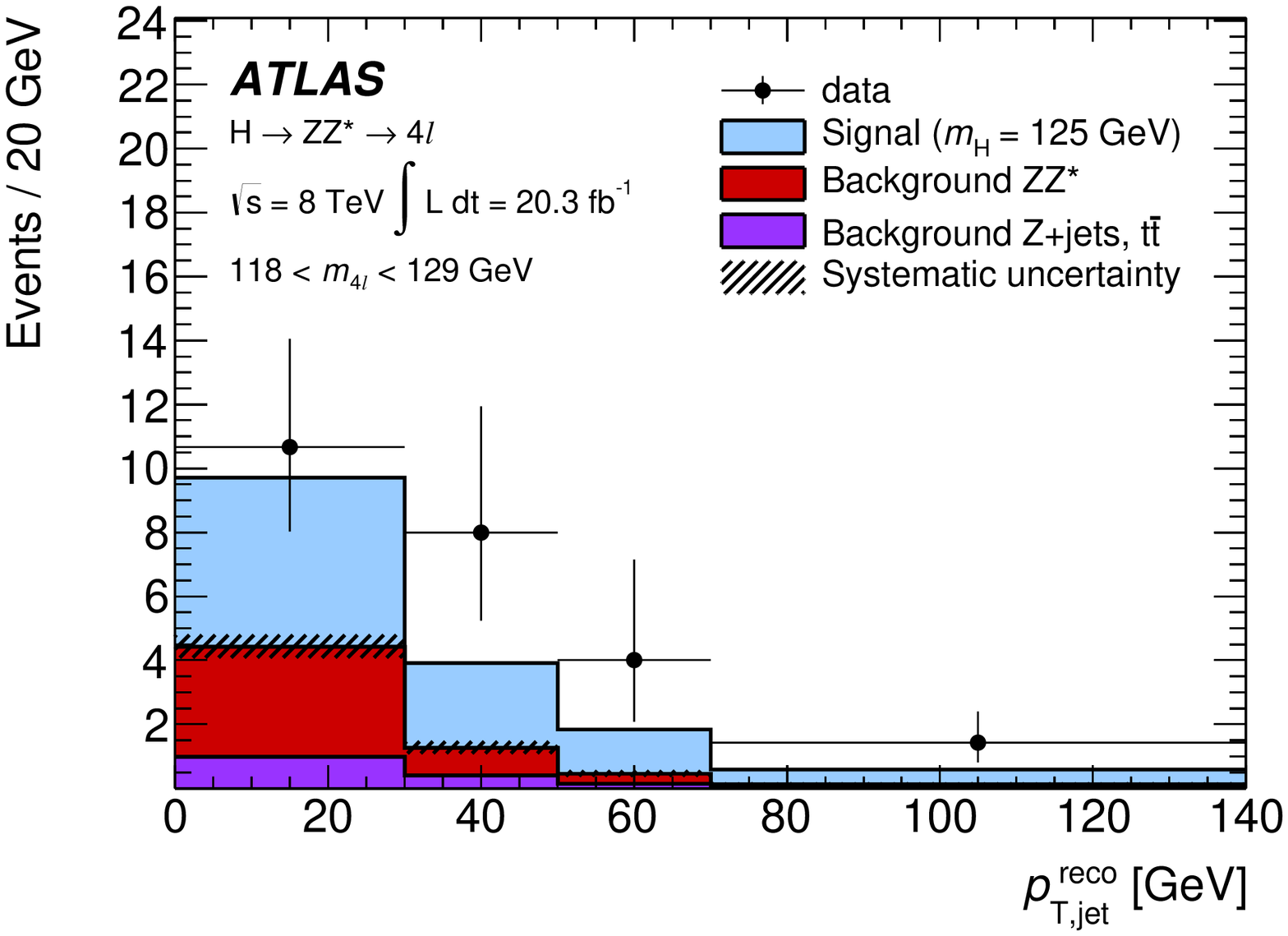}}
     
      \caption{\label{Fig:DiffObservedYieldsMain} Data yield distributions for the transverse momentum \ptHreco\ and the rapidity \yHreco\ of the four-lepton system, the invariant mass of the subleading lepton pair \mthreefourreco, the magnitude of the cosine of the decay angle of the leading lepton pair in the four-lepton rest frame with respect to the beam axis \costhetastarreco, the number of jets \njetsreco, and the transverse momentum of the leading jet \ptjetreco\ compared to signal and background expectations. The signal prediction includes VBF, $ZH$, $WH$, $t\bar{t}H$, and the \Powheg\ ggF calculation for a Higgs boson with $m_H$=125~GeV and is normalized to the most precise SM inclusive cross-section calculation currently available~\cite{Heinemeyer:2013tqa}. 
The hatched areas denote the systematic uncertainties on the backgrounds.}
 
     \end{center}
 \end{figure*}

\section{Observed differential yields and unfolding}
\label{sec:Unfolding}
The extraction of the signal yield for the measurement of the fiducial cross section is performed through a fit to the $m_{4\ell}$ distribution using 
shape templates for the signal and background contributions \cite{MassMuZZ}. In this fit, the Higgs boson mass is fixed to 125.4~GeV and the parameter of interest is the total number of signal events.
The extracted number of observed signal events in the mass window is 
23.7$^{+5.9}_{-5.3}$(stat.)$\pm 0.6$(syst.).  

In the differential cross-section measurements, given the low number of signal events expected in each measured bin $i$, the signal yields $\Nsigi$ 
are determined by subtracting the expected number of background events from the observed number of events. This is done within the mass window for each bin of the observable of interest. 
The total number of observed events in the mass window is 34 and the extracted signal yield is  
25.1$^{+6.3}_{-5.4}$(stat.)$^{+0.6}_{-0.4}$(syst.) events.

The difference between the number of signal events extracted with the two methods is mainly due to fixing the Higgs boson mass to 125.4~GeV in the fit method. As reported in Ref.~\cite{CombinedMass}, the best fit mass in the \HfourL\ channel alone is 124.5~GeV, causing smaller weights for some events in the fit.

After subtracting the background, the measured signal yields are corrected for detector efficiency and resolution effects.
This unfolding is performed using correction factors derived from simulated SM signal samples. 
The correction factor in the $i$-th bin is calculated as
\begin{equation*}
c_{i} = \frac{ N_i^{\mathrm{reco}}}{N^{\mathrm{fid}}_i} ,
\label{Eqn:CorrectionFactor}
\end{equation*}
where $N_i^{\mathrm{reco}}$ is the number of reconstructed events in the $i$-th bin of the observed distribution and $N^{\mathrm{fid}}_i$ is the number of events in the $i$-th bin of the particle-level distribution, within the fiducial region.
 
The unfolded signal yield in each bin is then converted into a differential fiducial cross section via
\begin{equation*}
   \dSigFidi  = \frac{\Nsigi}{ c_i \cdot \Lumi \cdot \Delta x_i},
   \label{eqn:FiducialCrossSection}
\end{equation*}
where $\Delta x_i$ is the bin width and $\Lumi$ the integrated luminosity.

The correction factors used in this analysis are obtained from simulated samples for all SM Higgs production modes, using the relative rates as predicted by the SM.
The inclusive correction factor is 
$c$~=~0.553~$\pm$~0.002(stat.)~$\pm$~0.015(syst.).
The correction factors for the different production modes are 
0.553 (ggF), 0.572 (VBF), 0.535 ($WH$), 0.551  ($ZH$) and 0.417 ($t\bar{t}H$).
In $t\bar{t}H$ production the Higgs boson is accompanied by light- and heavy-flavour jets as well as possible additional leptons from the top-quark decays. 
Since lepton isolation is applied to the reconstructed but not the fiducial objects, the correction factors for $t\bar{t}H$ differ from those for the other production modes.

For each bin, the number of expected background events, the number of observed events, the luminosity, and the correction factors are used to calculate a profile likelihood ratio~\cite{Cowan:2010js}. 
The likelihood includes shape and normalization uncertainties of backgrounds and correction factors as nuisance parameters.
For each variable all bins are included in the likelihood and correlations of uncertainties between the different bins and between backgrounds and correction factors are taken into account.
The cross sections are extracted for each bin by minimizing twice the negative logarithm of the profile likelihood ratio $-2 \ln{\Lambda}$. 
The uncertainties on the cross sections are also estimated using $-2 \ln{\Lambda}$ by evaluating its variation as a function of the parameter of interest (the cross section value in each bin).
Under the asymptotic assumption~\cite{Cowan:2010js},  $-2 \ln{\Lambda}$ behaves as a $\chi^2$ distribution with one degree of freedom. For some of the fitted intervals, due to the low number of events, the distribution of the profile likelihood ratio does not follow a $\chi ^2$ distribution and the uncertainties are derived using pseudo-experiments.

The compatibility between the measured cross sections and the theoretical predictions is evaluated by computing the difference between the value of $-2 \ln{\Lambda}$ at the best-fit value and the value obtained by fixing the cross sections in all bins to the ones predicted by theory.
Under the asymptotic assumption~\cite{Cowan:2010js}, this statistical observable behaves as a $\chi^2$ with the number of degrees of freedom equal to the number of bins;  it is used as a test statistic to compute the $p$-values quantifying the compatibility between the observed distributions and the predictions.
For all measured observables the asymptotic assumption is verified with pseudo-experiments.

\section{Systematic uncertainties}
\label{sec:Uncertainties}
 \begin{figure*}[htbp]
   \begin{center}
      \subfigure[\label{Fig:XS:pt}]{\includegraphics[width=0.43\linewidth]{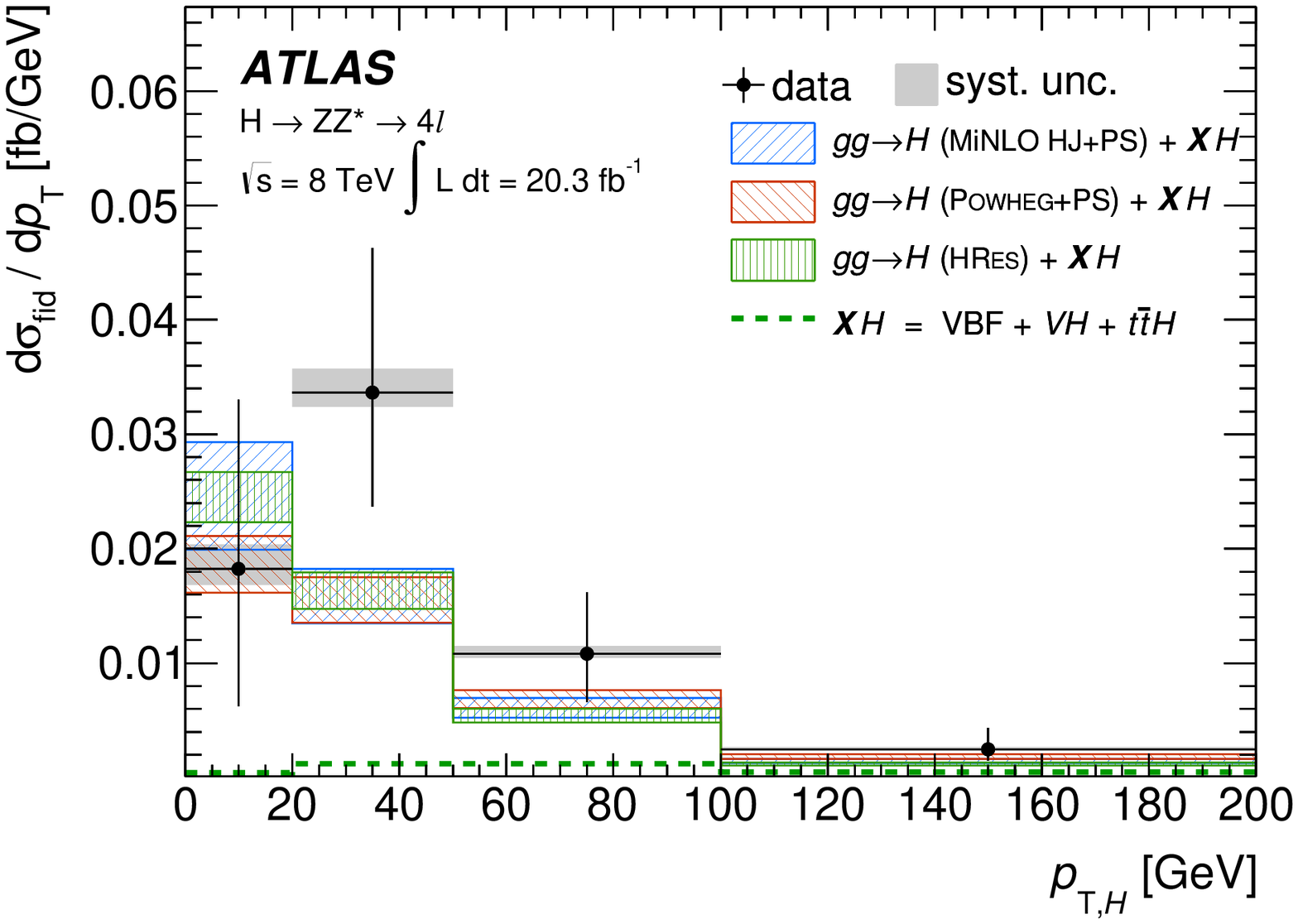}}
      \subfigure[\label{Fig:XS:rap}]{\includegraphics[width=0.43\linewidth]{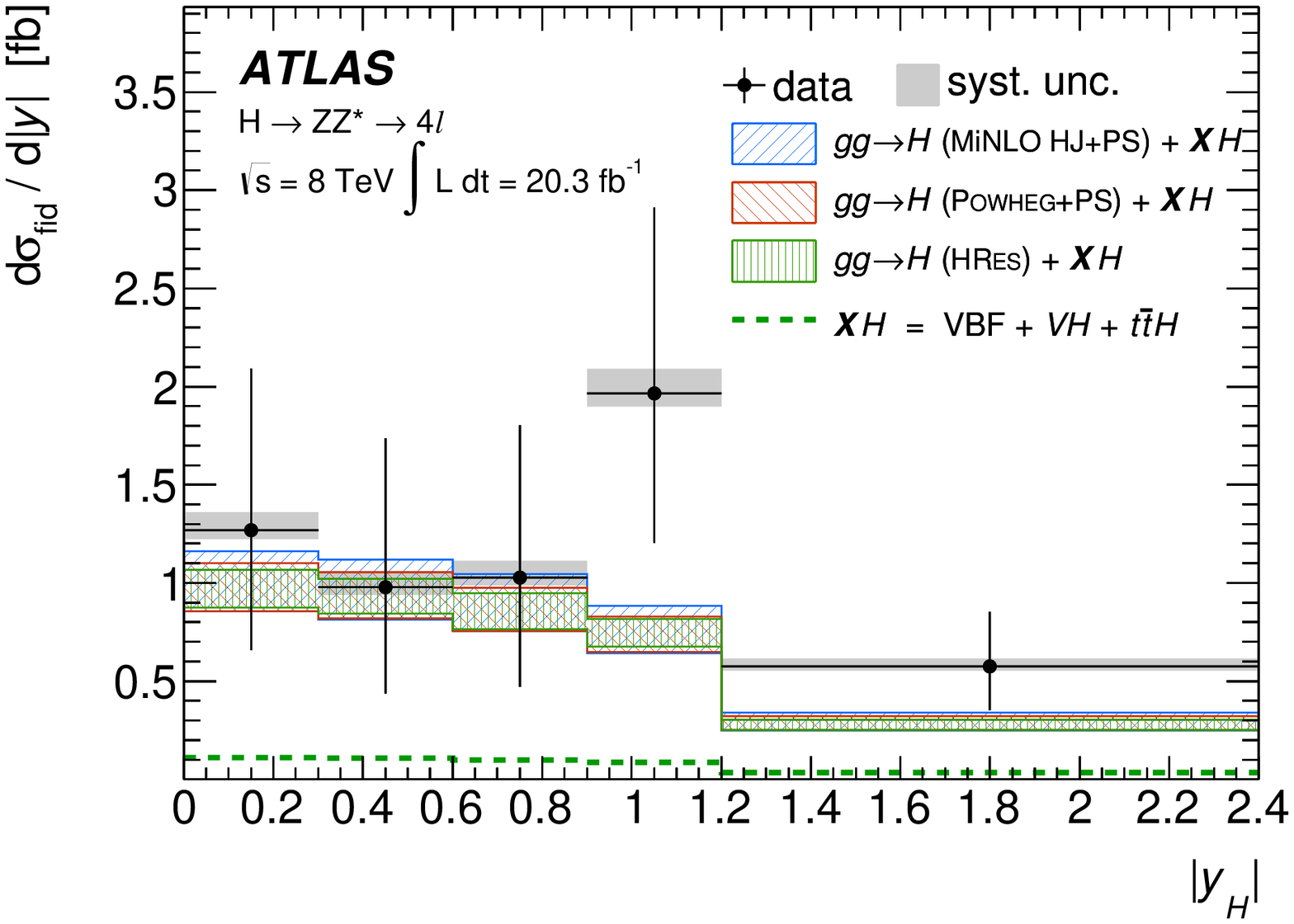}}
      \subfigure[\label{Fig:XS:m34}]{\includegraphics[width=0.43\linewidth]{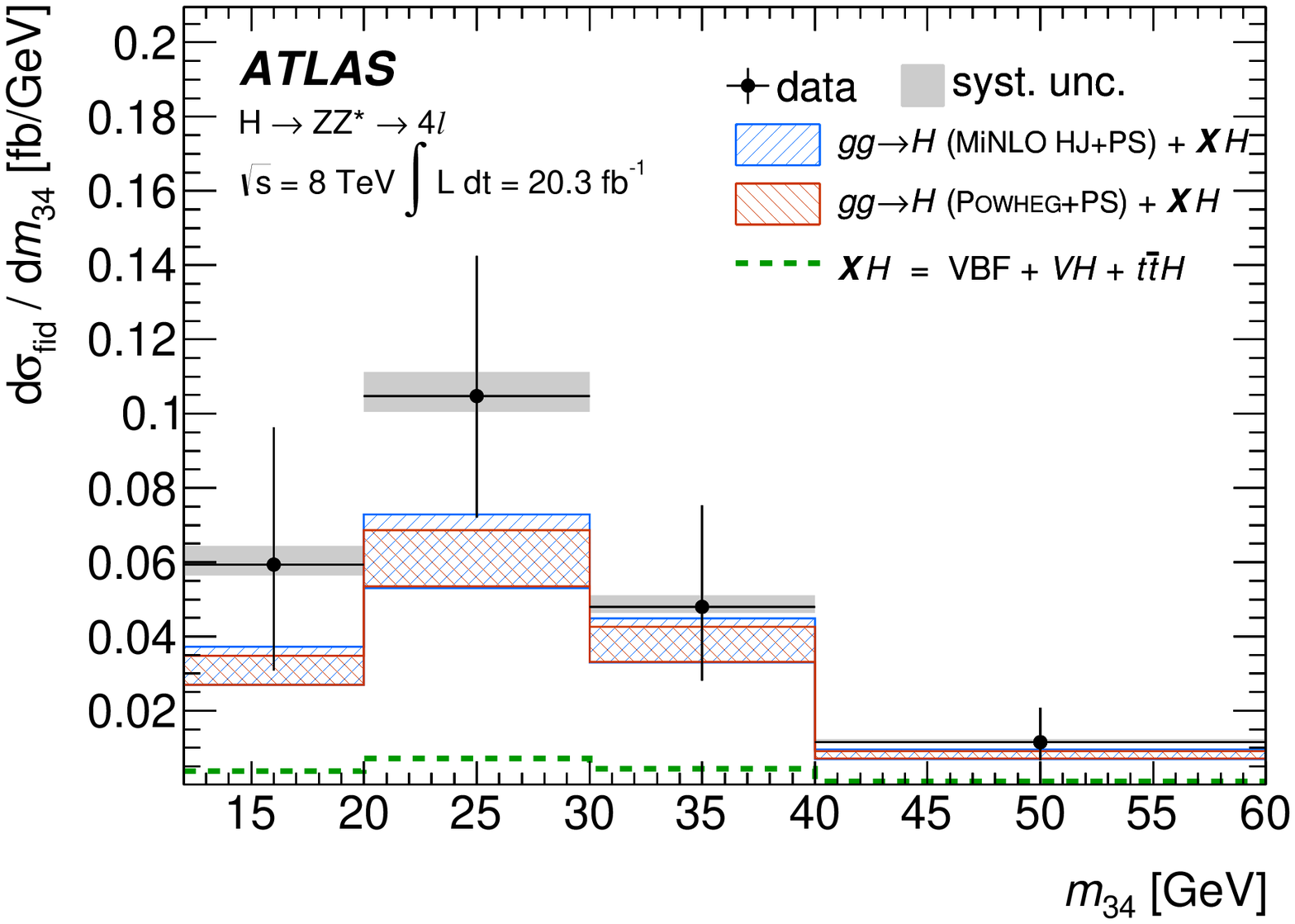}       }
     \subfigure[\label{Fig:XS:cos}]{\includegraphics[width=0.43\linewidth]{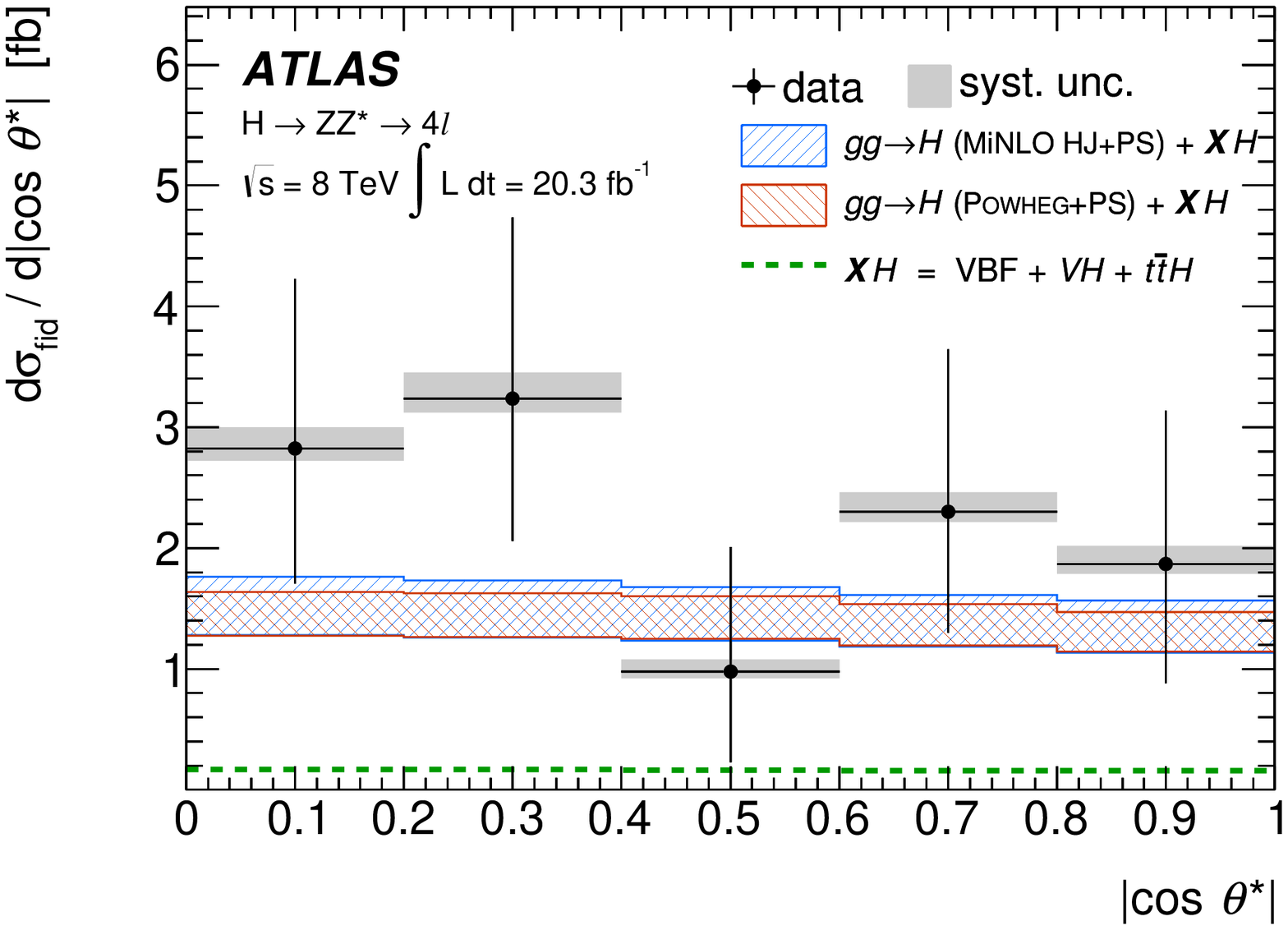} } 
      \subfigure[\label{Fig:XS:njets}]{\includegraphics[width=0.43\linewidth]{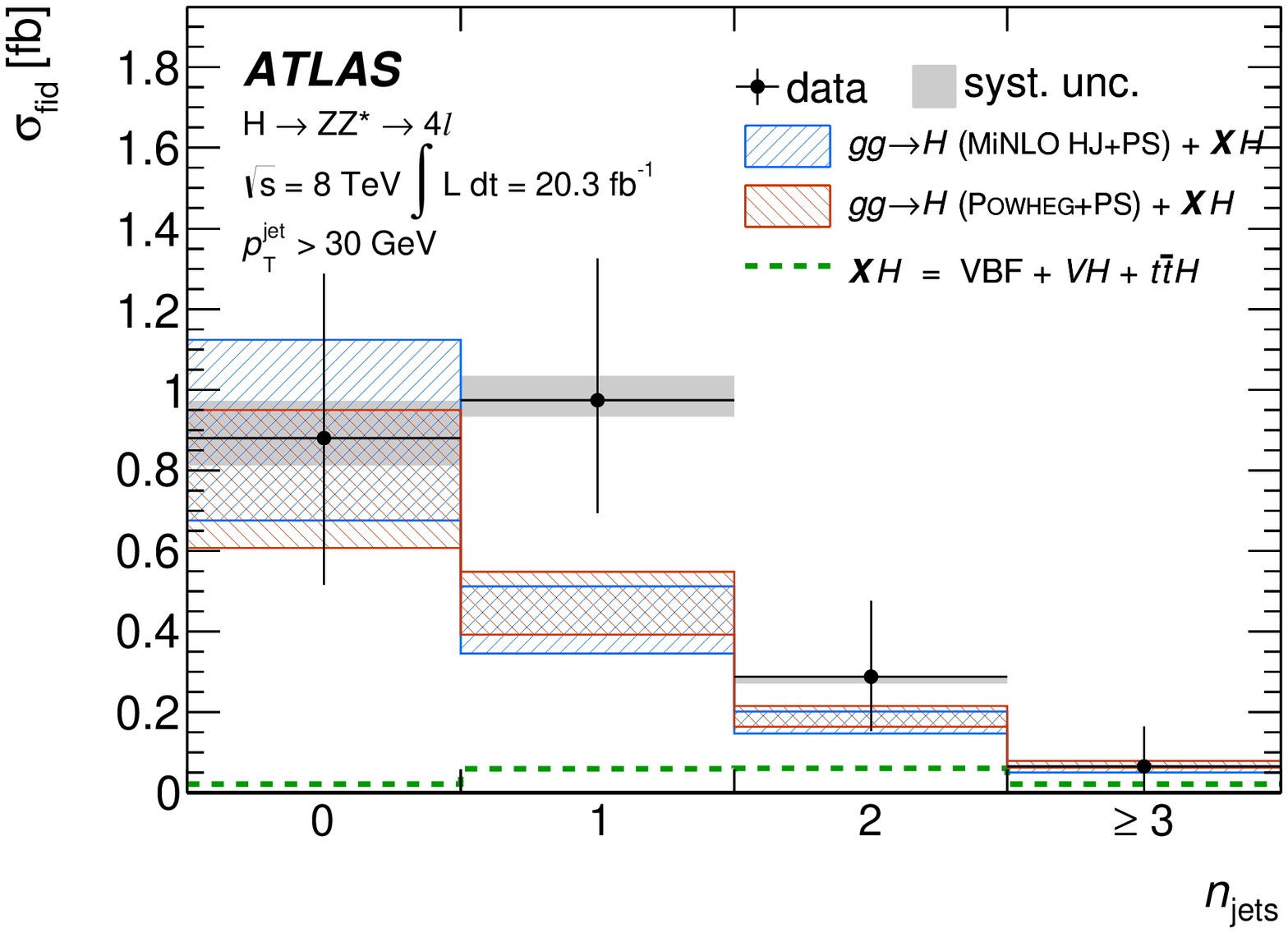} }
     \subfigure[\label{Fig:XS:jetpt}]{\includegraphics[width=0.43\linewidth]{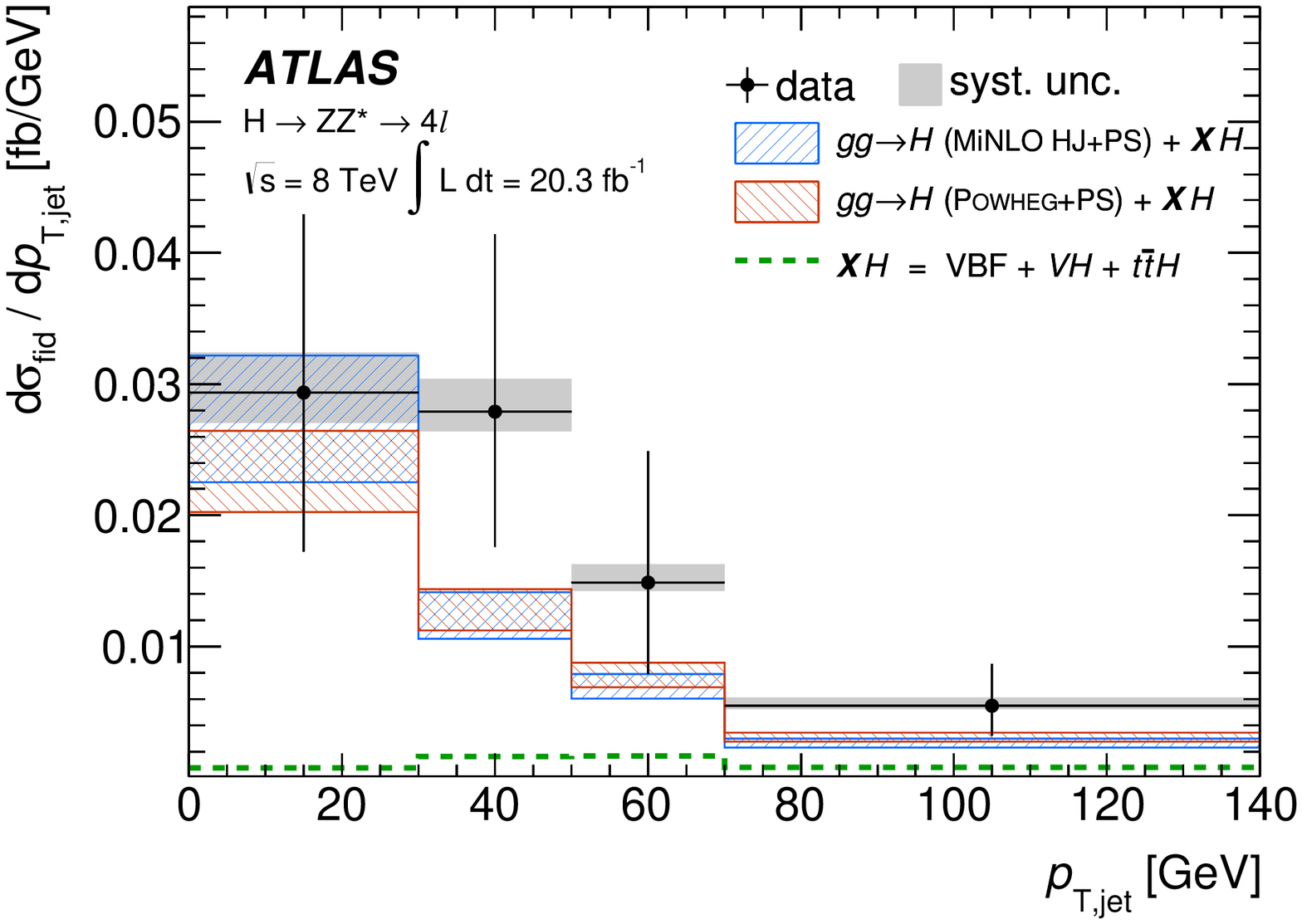}}
\caption{Differential unfolded cross sections for the transverse momentum \ptH\ and rapidity $y_H$ of the Higgs boson, the invariant mass of the subleading lepton pair $m_{34}$, the magnitude of the cosine of the decay angle of the leading lepton pair in the four-lepton rest frame with respect to the beam axis \costhetastar, the number of jets \njets, and the transverse momentum of the leading jet \ptjet\ in the \HfourL\ decay channel compared to different theoretical calculations of the ggF process: \Powheg, \Minlo\ and \HResTwo. The contributions from VBF, $ZH/WH$ and $t\bar{t}H$ are determined as described in Section~\ref{sec:Samples} and added to the ggF distributions. All theoretical calculations are normalized to the most precise SM inclusive cross-section predictions currently available~\cite{Heinemeyer:2013tqa}. The error bars on the data points show the total ($\mathrm{stat.} \oplus \mathrm{syst.}$) uncertainty, while the grey bands denote the systematic uncertainties. The bands of the theoretical prediction indicate the total uncertainty. 
\label{fig:XS}}
    \end{center}
\end{figure*}
Systematic uncertainties are calculated for the estimated backgrounds, the correction factors, 
and the SM theoretical predictions; the latter only have an impact on the quantitative comparison of the measurements with different predictions.
An overview of the systematic uncertainties on the total background prediction and the correction factors is shown in Table~\ref{tab:systs}.
\begin{table}
\centering
\caption{\label{tab:systs} Summary of the relative systematic uncertainties on the total background contribution (top rows) and on the parameters that enter the signal extraction (bottom rows). 
The ranges indicate the variation across observables and bins.}
\begin{tabular}{ll}
\hline
\hline
\multicolumn{2}{c}{Systematic Uncertainties (\%)} \tabularnewline
\hline
  \multicolumn{2}{l}{\em Background} \tabularnewline
      \hspace{.4cm} Luminosity               & $1.4$--$2.3$  \\ 
      \hspace{.4cm} Reducible background & $1.6$--$34$    \\
      \hspace{.4cm} Experimental, leptons             & $1.3$--$2.3$ \\
      \hspace{.4cm} PDF/scale        & $3.0$--$24$ \tabularnewline
  \multicolumn{2}{l}{\em Correction factors/conversion to $\sigma$} \\
      \hspace{.4cm} Luminosity               & $2.8$  \\
      \hspace{.4cm} Experimental, leptons       & $2.1$--$2.6$\\
      \hspace{.4cm} Experimental, jets        & $2.7$--$13$\\
      \hspace{.4cm} Production process  & $0.1$--$15$ \\
      \hspace{.4cm} Higgs boson mass                & $0.4$--$2.7$ \\
\hline
\hline
\end{tabular}
\end{table}

The uncertainty on the integrated luminosity is propagated in a correlated way to the backgrounds evaluated from the MC predictions and to the unfolding, 
where it is used when converting the estimated unfolded signal yield into a fiducial cross section. This uncertainty is derived following the same methodology as 
that detailed in Ref.~\cite{Aad:2013ucp} from a preliminary calibration of the luminosity scale derived from beam-separation scans performed in November 2012.

Systematic uncertainties on the data-driven estimate of the reducible backgrounds are assigned both to the normalization and the shapes of the distributions by varying the estimation methods \cite{MassMuZZ}. 

The systematic uncertainties on the lepton trigger, reconstruction and identification efficiencies \cite{ElectronEff2012,muonpaper} are propagated to the signal correction factors and the $ZZ^*$ background, taking into account correlations.
For the correction factors, systematic uncertainties are assigned on the jet resolution and energy scales.
The largest systematic uncertainty is due to the uncertainty in the jet flavour composition \cite{jetcalib,JER,JES}.

The uncertainties on the correction factors due to PDF choice as well as QCD renormalization and factorization scale variations are evaluated in signal samples using the procedure described in Ref.~\cite{MassMuZZ} and found to be negligible.
A similar procedure is followed for most variables for the irreducible \ZZ\ background. For the jet-related observables an uncertainty is derived instead by comparing the data with the predicted \ZZ\ distributions for $m_{4\ell} >$~190~GeV, after normalizing the MC estimate to the observed data yield. The systematic uncertainty is estimated as the larger of the data-MC difference and the statistical uncertainty on the data. This systematic uncertainty accounts for both the theoretical and experimental uncertainties in the modelling of the \ZZ\ jet distributions. Systematic uncertainties due to the modelling of QED final-state radiation are found to be negligible with respect to the total uncertainty. 

The correction factors are calculated  assuming the predicted relative cross sections of the different Higgs production modes.
The corresponding systematic uncertainty is evaluated by varying these predictions within the current experimental bounds \cite{Aad:2013wqa}. The VBF and $VH$ fractions are varied by factors of 0.5 and 2 with respect to the SM prediction and the $t\bar{t}H$ fraction is varied by factors of 0 and 5.

The experimental uncertainty on $m_H$ \cite{CombinedMass} is propagated to the correction factors by studying their dependence on the Higgs boson mass.

The systematic uncertainties on the theoretical predictions include the PDF and QCD scale choices as well as the uncertainty on the $H \rightarrow\ ZZ^*$ branching fraction \cite{Heinemeyer:2013tqa}. The procedure described in Ref.~\cite{Stewart:2011cf} is used to evaluate the scale uncertainties of the predicted \njets\ distribution.

The upper edges of the uncertainty ranges in Table~\ref{tab:systs} are in most cases due to the highest bins in the \njets\ and \ptjet\ distributions. 
The background systematic uncertainties are large in some bins due to the limited statistics in the data control regions.

\section{Results}
\label{sec:Results}
The cross section in the fiducial region described in Table~\ref{tab:FiducialRegion} is 
\begin{equation*}
\sigma^{\mathrm{fid}}_{\mathrm{tot}} =2.11^{+0.53}_{-0.47}\mathrm{(stat.)} \pm 0.08 \mathrm{(syst.)~\mathrm{fb}}.
\end{equation*}
The theoretical prediction from Ref.~\cite{Heinemeyer:2013tqa} for a Higgs boson mass of 125.4 GeV is $1.30 \pm 0.13$~fb.

The differential cross sections as a function of \ptH, $y_H$, $m_{34}$, \costhetastar, \njets, and \ptjet\ are shown in Fig.~\ref{fig:XS}.   
For all variables and bins the total uncertainties on the cross-section measurements are dominated by statistical uncertainties.
\Powheg, \Minlo\ and \HResTwo\ calculations of ggF, added to VBF, $ZH/WH$ and $t\bar{t}H$ (see Section~\ref{sec:Samples}), are overlaid. 
The \HResTwo\ calculation was developed for modelling the Higgs kinematic variables and is only used for \ptH\ and $y_H$.
The theoretical calculations are normalized to the most precise SM inclusive cross-section predictions currently available~\cite{Heinemeyer:2013tqa}.

The $p$-values quantifying the compatibility between data and predictions, computed with the method described in Section~\ref{sec:Unfolding}, 
are shown in Table~\ref{Tab:Compatibility}. No significant discrepancy is observed.

\begin{table}
\centering
\caption{\label{Tab:Compatibility}
Compatibility tests of data with \Powheg, \Minlo\ and \HResTwo\ ggF calculations of SM Higgs boson production.  
The compatibility $p$-values are obtained, as explained in the text, from the difference between $-2 \ln{\Lambda}$ at the best-fit value and  $-2 \ln{\Lambda}$ with the cross sections fixed to the theory computations.}
\begin{tabular}{lccc}
\hline
\hline
 		       	& \multicolumn{3}{c}{  $p$-values } 	\\
 Variable       	& \Powheg   & \Minlo & \HResTwo  \\
\hline                                                    
\ptH 			& 0.30 	& 0.23 &  0.16 \\
$|y_H|$ 		& 0.37 	& 0.45 &  0.36 \\
$m_{34}$ 	        & 0.48 	& 0.60 &  -\\
\costhetastar 	        & 0.35 	& 0.45 &  -\\
\njets	                & 0.37	& 0.28 &  -\\
\ptjet 		        & 0.33	& 0.26 &  -\\
\hline
\hline
\end{tabular}
\end{table}

\section{Conclusion}
\label{sec:Conclusion}
Measurements of fiducial and differential cross sections in the \HfourL\ decay channel are presented. 
They are based on 20.3~\ifb\ of $pp$ collision data, produced at $\sqrt{s}$~=~8~TeV centre-of-mass energy at the LHC 
and recorded by the ATLAS detector. 
The cross sections are corrected for detector effects and compared to selected theoretical calculations. 
No significant deviation from the theoretical predictions is observed for any of the studied variables.

\section*{Acknowledgements}
We thank CERN for the very successful operation of the LHC, as well as the
support staff from our institutions without whom ATLAS could not be
operated efficiently.

We acknowledge the support of ANPCyT, Argentina; YerPhI, Armenia; ARC,
Australia; BMWF and FWF, Austria; ANAS, Azerbaijan; SSTC, Belarus; CNPq and FAPESP,
Brazil; NSERC, NRC and CFI, Canada; CERN; CONICYT, Chile; CAS, MOST and NSFC,
China; COLCIENCIAS, Colombia; MSMT CR, MPO CR and VSC CR, Czech Republic;
DNRF, DNSRC and Lundbeck Foundation, Denmark; EPLANET, ERC and NSRF, European Union;
IN2P3-CNRS, CEA-DSM/IRFU, France; GNSF, Georgia; BMBF, DFG, HGF, MPG and AvH
Foundation, Germany; GSRT and NSRF, Greece; ISF, MINERVA, GIF, DIP and Benoziyo Center,
Israel; INFN, Italy; MEXT and JSPS, Japan; CNRST, Morocco; FOM and NWO,
Netherlands; BRF and RCN, Norway; MNiSW, Poland; GRICES and FCT, Portugal; MERYS
(MECTS), Romania; MES of Russia and ROSATOM, Russian Federation; JINR; MSTD,
Serbia; MSSR, Slovakia; ARRS and MIZ\v{S}, Slovenia; DST/NRF, South Africa;
MICINN, Spain; SRC and Wallenberg Foundation, Sweden; SER, SNSF and Cantons of
Bern and Geneva, Switzerland; NSC, Taiwan; TAEK, Turkey; STFC, the Royal
Society and Leverhulme Trust, United Kingdom; DOE and NSF, United States of
America.

The crucial computing support from all WLCG partners is acknowledged
gratefully, in particular from CERN and the ATLAS Tier-1 facilities at
TRIUMF (Canada), NDGF (Denmark, Norway, Sweden), CC-IN2P3 (France),
KIT/GridKA (Germany), INFN-CNAF (Italy), NL-T1 (Netherlands), PIC (Spain),
ASGC (Taiwan), RAL (UK) and BNL (USA) and in the Tier-2 facilities
worldwide.

\bibliographystyle{elsarticle-num}
\bibliography{H4lpaper}

\onecolumn 
\clearpage
\begin{flushleft}
{\Large The ATLAS Collaboration}

\bigskip

G.~Aad$^{\rm 85}$,
B.~Abbott$^{\rm 113}$,
J.~Abdallah$^{\rm 153}$,
S.~Abdel~Khalek$^{\rm 117}$,
O.~Abdinov$^{\rm 11}$,
R.~Aben$^{\rm 107}$,
B.~Abi$^{\rm 114}$,
M.~Abolins$^{\rm 90}$,
O.S.~AbouZeid$^{\rm 160}$,
H.~Abramowicz$^{\rm 155}$
% ...
\bigskip
\\
% ...
$^{11}$ Institute of Physics, Azerbaijan Academy of Sciences, Baku, Azerbaijan\\
% ...
$^{85}$ CPPM, Aix-Marseille Universit{\'e} and CNRS/IN2P3, Marseille, France\\
% ...
$^{90}$ Department of Physics and Astronomy, Michigan State University, East Lansing MI, United States of America\\
% ...
$^{107}$ Nikhef National Institute for Subatomic Physics and University of Amsterdam, Amsterdam, Netherlands\\
% ...
$^{113}$ Homer L. Dodge Department of Physics and Astronomy, University of Oklahoma, Norman OK, United States of America\\
$^{114}$ Department of Physics, Oklahoma State University, Stillwater OK, United States of America\\
% ...
$^{117}$ LAL, Universit{\'e} Paris-Sud and CNRS/IN2P3, Orsay, France\\
% ...
$^{153}$ Institute of Physics, Academia Sinica, Taipei, Taiwan\\
% ...
$^{155}$ Raymond and Beverly Sackler School of Physics and Astronomy, Tel Aviv University, Tel Aviv, Israel\\
% ...
$^{160}$ Department of Physics, University of Toronto, Toronto ON, Canada\\
% ...
\end{flushleft}


\end{document}